\documentclass[superscriptaddress,pre,twocolumn,preprintnumbers,amsmath,amssymb]{revtex4-1}
\usepackage{epsfig}
\usepackage{amsmath}
\usepackage{array}
\usepackage{multirow}
\usepackage{psfrag}
\usepackage[font=footnotesize,format=plain,labelsep=period,justification=raggedright]{caption}
\usepackage{subcaption}
\usepackage{graphicx}
\usepackage{color}
\usepackage{comment}
\usepackage{float}

\begin{document}
\title{Capillary Bridges on Liquid Infused Surfaces}
\author{Alvin C. M. Shek}
\affiliation{Department of Physics, Durham University, Durham, DH1 3LE, UK}
\author{Ciro Semprebon}
\affiliation{Department of Mathematics, Physics and Electrical Engineering, 
Northumbria University, Newcastle upon Tyne NE1 8ST, UK}
\author{Jack R. Panter}
\affiliation{Department of Physics, Durham University, Durham, DH1 3LE, UK}
\author{Halim Kusumaatmaja}
\email{Email: halim.kusumaatmaja@durham.ac.uk}
\affiliation{Department of Physics, Durham University, Durham, DH1 3LE, UK}
\date{\today}
	
\begin{abstract}
We numerically study two-component capillary bridges formed when a liquid droplet is placed in between two liquid infused surfaces (LIS). In contrast to commonly studied one-component capillary bridges on non-infused solid surfaces, two-component liquid bridges can exhibit a range of different morphologies where the liquid droplet is directly in contact with two, one or none of the LIS substrates. In addition, the capillary bridges may lose stability when compressed due to the envelopment of the droplet by the lubricant. We also characterise the capillary force, maximum separation and effective spring force, and find they are influenced by the shape and size of the lubricant ridge. Importantly, these can be tuned to increase the effective capillary adhesion strength by manipulating the lubricant pressure, Neumann angle, and wetting contact angles. As such, LIS are not only ``slippery'' parallel to the surface, but they are also ``sticky'' perpendicular to the surface.
\end{abstract}
	
	\maketitle
	\section{Introduction}
	
	The subject of capillary bridges has received much attention in the literature due to their ubiquity in nature and engineering applications.  For example, capillary bridges between parallel flat plates \cite{carter_forces_1988,klingner2004capillary,broesch2012concave} have been considered for applications in liquid transfer \cite{kang2008liquid,dodds2011stretching}, wet adhesion device \cite{lee2007reversible,vogel2010capillarity}, and for the formation of curved polymeric particles \cite{wang2015capillary}, while those between curved solid bodies \cite{kohonen2004capillary,de1999particle,pakarinen2005towards} occur in wet granular materials \cite{mitarai2006wet,scheel2008morphological} and in atomic force microscopy experiments \cite{xu1998wetting,xiao2000investigation,wei2007growth}. Capillary bridges also play important roles in the physiology of numerous insects and animals, such as in the adhesive pads of Asian Weaver ants \cite{beutel2001ultrastructure,federle2001biomechanics} and how shorebirds trap and consume prey inside their beaks \cite{prakash2008surface}. Furthermore, the stability of liquid bridge shapes and their force-separation relations have been investigated for smooth and patterned surfaces \cite{de2008effect,nosonovsky2008capillary,zhuang2015combined,banerjee2012effect}, as well as in the presence of one and multiple droplets \cite{banerjee2012effect,gui1995stiction,de2008enhancement}. 
	
	In contrast to previous works, where the capillary bridges comprise of a single liquid component, here we study the case where they have two liquid components, especially in the context of the so-called liquid infused surfaces (LIS) or slippery liquid infused porous surfaces (SLIPS). These are a novel class of functional surfaces constructed by infusing textured or porous materials with wetting lubricants \cite{lafuma_slippery_2011,smith2013droplet,wong2011bioinspired}. Many applications of LIS exploit the fact that, on LIS, liquids move easily parallel to the substrate \cite{lafuma_slippery_2011,smith2013droplet,wong2011bioinspired,kim2012liquid,anand2012enhanced,wexler2015shear,villegas2019liquid}. This ``slippery'' nature results in numerous advantageous properties, such as self-cleaning, enhanced heat transfer, anti-fouling, and anti-icing. Here, we focus instead on the adhesive properties of droplets when displaced perpendicularly to the substrate. Since the adhesive force now has contributions which originate not just from the liquid droplet but also from the lubricant, we will argue that LIS are ``sticky'' in the perpendicular direction.

\begin{figure}
	\centering
	\includegraphics[width=\linewidth]{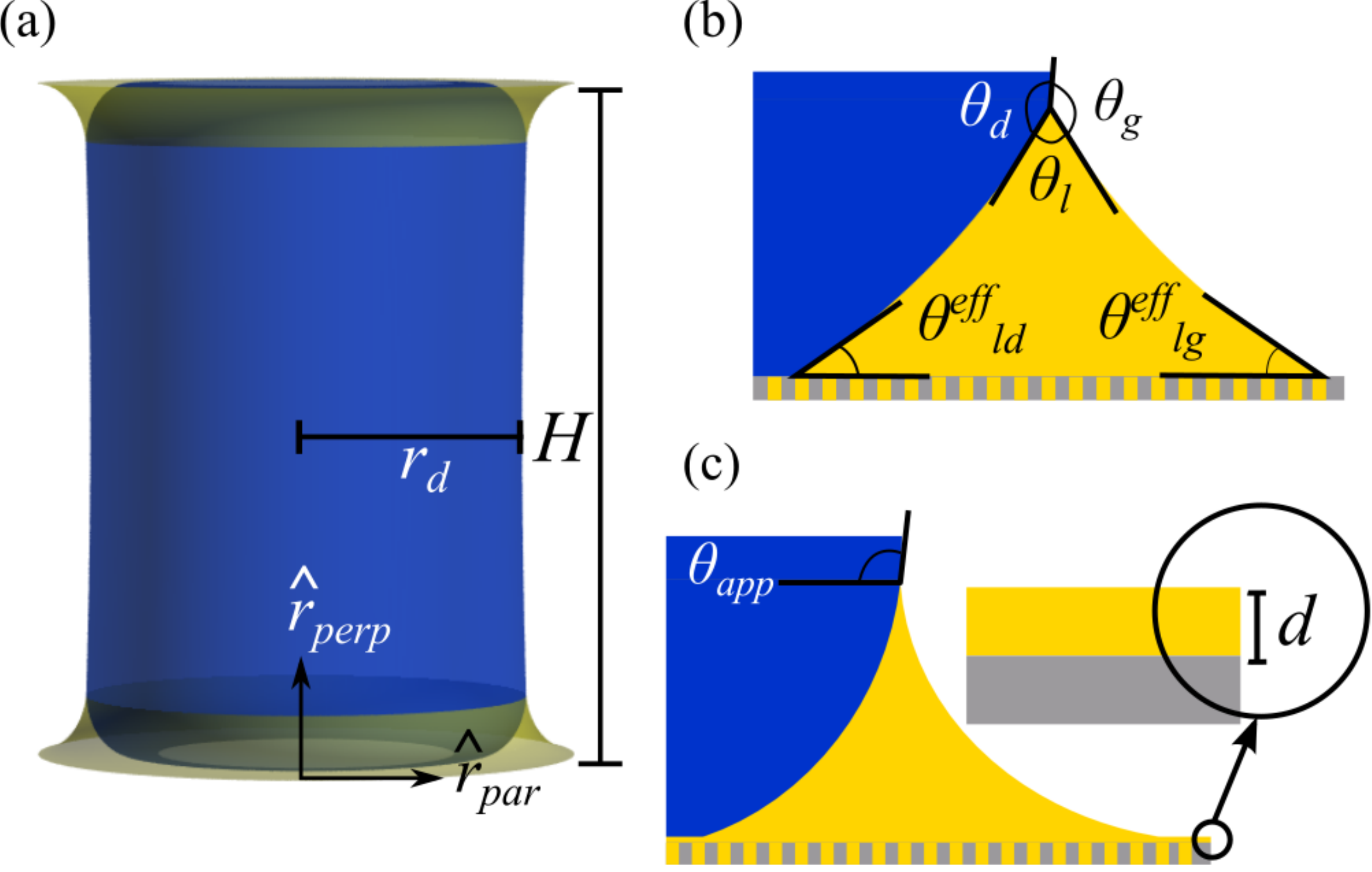}
	\caption{Capillary bridge geometry and parameters. (a) Capillary bridge between two LIS, where $H$ is the separation distance, and $\hat{r}_{perp}$, $\hat{r}_{par}$ are the unit vectors perpendicular and parallel to the substrate. Panels (b) and (c) show the capillary bridge cases where the lubricant partially and completely wetting the solid corrugations, respectively.  
	Additionally, panel (b) illustrates the definition of the Neumann angles, $\theta_{l}$, $\theta_{d}$, $\theta_{g}$, and the effective wetting contact angles, $\theta^{\rm eff}_{ld}$ and $\theta^{\rm eff}_{lg}$. Panel (c) also gives a geometrical interpretation of the apparent contact angle, $\theta_{app}$, and the lubricant film thickness, $d$. 
	\label{LIS_3DParameters}}
	\end{figure}
	
	To study capillary adhesion on LIS, we will consider a liquid droplet sandwiched between two parallel LIS, as illustrated in Fig. \ref{LIS_3DParameters} (a). For simplicity, we assume that gravity is negligible and consider symmetric surface properties such that the Neumann ($\theta_{l}$, $\theta_{d}$ and $\theta_{g}$) and wetting contact angles ($\theta^{\rm eff}_{ld}$ and $\theta^{\rm eff}_{lg}$), illustrated in Fig. \ref{LIS_3DParameters} (b) and (c), will be the same for the top and bottom plates. In this work, we focus on equilibrium rather than dynamic effects, obtained by quasi-statically varying the separation between the substrates. 
	
	The fact that capillary bridges on LIS involve two liquid components, instead of just one, leads to several interesting phenomena. Firstly, two-component liquid bridges can exhibit a wider range of interfacial morphologies and topologies. Morphological transitions can be triggered by compressing and stretching the capillary bridges. Secondly, to characterise the ``stickiness'' of LIS, we calculate the capillary force and effective spring constant of the liquid bridges. We find both the size and shape of the lubricant ridge to play an important role. In the limit of vanishing ridge size, we observe that the adhesion force converges to that of a perfectly smooth surface. Since perfectly smooth surfaces are challenging to realise, LIS can be considered as an excellent alternative to mimic their properties in the limit of small ridge. Additionally, LIS can also be tuned to have stronger adhesion than an equivalent smooth surface. 
	
	This paper is organized as follows. In Section~\ref{model} we introduce our theoretical framework to study the capillary bridges.  
Then, in Section~\ref{results}, we organize our results in terms of the morphology diagrams, capillary forces and spring constants of the capillary bridges as function of the relevant control parameters. Finally, we summarize our results in the Conclusion Section.

	\section{Model and Method} \label{model}
	This section is divided into three parts. In the first part we introduce the model for capillary bridges between two LIS. Then, we describe our numerical scheme to compute the equilibrium bridge morphologies. Finally, we provide the expressions for calculating the capillary force and spring constant.

	\subsection{Model}
	We begin by considering the case of a one-component liquid bridge. The total free energy \cite{brinkmann_wetting_2002,kusumaatmaja2010equilibrium} is given by 
	\begin{eqnarray}
	E_{\rm{smooth}} &=& \gamma_{dg}A_{dg} + \gamma_{ds}A_{ds} + \gamma_{gs}A_{gs} - \Delta P_{dg} V_{d}, \\
	&=& \gamma_{dg}A_{dg} - \gamma_{dg} \cos \theta_{dg} A_{ds} - \Delta P_{dg} V_{d} + {\rm{C}_1}, \nonumber
	\end{eqnarray}
	where the subscripts $s$, $d$, and $g$ stand for solid, drop and gas respectively; $\gamma_{ij}$ and $A_{ij}$ are the surface tension and interfacial area between components $i$ and $j$, $V_{d}$ is the drop volume; and $\Delta P_{dg}$ is the pressure difference between the drop and gas components. We assume the liquid drop is non-volatile, so that its volume is a conserved quantity.  For a fixed drop volume $V_{d}$, the term $\Delta P_{dg}$ can be interpreted as a Lagrange multiplier. Rearranging the terms, and assuming the capillary bridge is confined by ideal and smooth surfaces, the liquid wettability is expressed by the material contact angle, $\cos \theta_{dg} = (\gamma_{gs} - \gamma_{ds}) / \gamma_{dg}$. The remaining terms sum up to a constant, ${\rm{C}_1} = \gamma_{gs} (A_{gs}+A_{ds})$, which do not alter the liquid bridge morphology, and can be neglected.
	
Now we turn our attention to a capillary bridge sandwiched between two LIS substrates, as depicted in Fig. \ref{LIS_3DParameters}(a). Here the lubricant ridges are formed around the droplet due to capillary action. As in our previous works  \cite{semprebon2017apparent,sadullah2018drop,sadullah2020bidirectional}, 
we divide the total free energy contributions into two parts, the fluid-fluid and fluid-solid contributions, such that $E_{\rm LIS} = E_{\rm FF} + E_{\rm FS}$. The fluid-fluid contributions are 
\begin{equation}
E_{\rm FF} = \gamma_{dg}A_{dg} + \gamma_{ld}A_{ld} + \gamma_{lg}A_{lg} - \Delta P_{dg} V_{d} - \Delta P_{lg} V_{l}. \label{eq:FF}
\end{equation}
Similar to the one-component liquid bridge case, the drop volume is assumed to be constant. For the lubricant (subscript $l$), we assume the rest of the substrate infused by the lubricant provides a virtually infinite reservoir. In experiments the lubricant exchange between the ridge and surrounding substrate can occur on a rather slow timescale due to the strong viscous dissipation in the thin lubricant layer \cite{kreder2018film}. However, we will assume this exchange is still much faster than the typical variation in the control parameters used in this work, such as the separation between the two LIS. Consequently, we employ the pressure ensemble for the lubricant, parameterised by the pressure jump $\Delta P_{lg}$ at the lubricant-gas interface. The term $-\Delta P_{lg} V_{l}$ in the free energy represents the energy cost for drawing additional lubricant from the reservoir.
	
When the lubricant only partially wets (PW) the solid surface, see Fig. \ref{LIS_3DParameters}(b), we can write the fluid-solid contributions as
\begin{eqnarray}
E_{\rm FS(PW)} & =& \gamma^{\rm eff}_{ds}A^{\rm pr}_{ds}+\gamma^{\rm eff}_{ls}A^{\rm pr}_{ls}+\gamma^{\rm eff}_{gs}A^{\rm pr}_{gs} \\
& = & \gamma_{ld}A^{\rm pr}_{ds}\cos\theta^{\rm eff}_{ld} - \gamma_{lg}(A^{\rm pr}_{ds} + A^{\rm pr}_{ls})\cos\theta^{\rm eff}_{lg} + {\rm{C_2}}, \nonumber
\end{eqnarray}
where $A^{\rm pr}_{is}$ is the projected interfacial area between fluid $i$ and the substrate. The index $i=d,g$ represents either the drop or gas phase. It is worth emphasizing that the substrate is effectively a composite of the underlying rough solid surface and the imbibed lubricant. For simplicity, we will not resolve the details of the composite surface. Instead, we simply assume this gives rise to an effective average surface tension $\gamma^{\rm eff}_{is} = f \gamma_{is} + (1-f) \gamma_{il}$, with $f$ the fraction of the projected solid area exposed to the drop or gas phase. We also define the effective contact angle $\cos\theta^{\rm eff}_{ij} = (\gamma^{\rm eff}_{js} - \gamma^{\rm eff}_{is})/\gamma_{ij}$ between phases $i$ and $j$ on the composite solid-lubricant substrate. The constant term ${\rm{C}_2} = \gamma^{\rm eff}_{gs}(A^{\rm pr}_{ds}+A^{\rm pr}_{ls}+A^{\rm pr}_{gs})$ can be neglected for the same reason that it will not affect the resulting morphology.
	
The limit of $\theta_{ld}$, $\theta_{lg}\rightarrow 0^\circ$ describes the complete wetting (CW) case, Fig. \ref{LIS_3DParameters}(c), where the lubricant will form a thin layer above the surface. The thickness of the lubricant layer is determined by the intermolecular interactions between lubricant and solid \cite{daniel2017oleoplaning}. These interactions are also often called the disjoining pressure term. To extend our calculations to this limit, we introduce
\begin{equation}
E_{\rm FS(CW)} = \int \frac{B}{12\pi d(\mathbf{r})^2} \text{ d}A,
\end{equation}
where $B$ is a constant and $d$ is the thickness of the lubricant layer. A different choice for this contribution would not alter the results of this study, as long as the film thickness is small compared to the size of both the drop and the lubricant ridge. In practice, we also do not observe major differences in the results for full wetting and small but finite contact angles.

On LIS, the drop-gas interface does not come in contact with the solid, due to the ubiquitous presence of a lubricant ridge, see Fig. \ref{LIS_3DParameters}.  At the top of this lubricant ridge, there is a triple contact line where the drop-gas (\textit{dg}) interface meets the lubricant-gas (\textit{lg}) and lubricant-drop (\textit{ld}) interfaces. The three Neumann angles, $\theta_{l}$, $\theta_{d}$ and $\theta_{g}$, are related to the interfacial tensions via
\begin{align}\label{NeumannAngle_Eqn}
\frac{\gamma_{ld}}{\sin \theta_{g}} = \frac{\gamma_{dg}}{\sin \theta_{l}} = \frac{\gamma_{lg}}{\sin \theta_{d}}.
\end{align}  
Consequently, on LIS the definition of a material contact angle needs to be adapted. Two alternatives are possible: either taking the slope of the drop-gas interface immediately above the triple line, or by estimating the slope of the (virtual) drop-gas interface if it were to continue within the lubricant ridge. Following Semprebon \textit{et al}. \cite{semprebon2017apparent}, we employ the first definition, as illustrated in Fig. \ref{LIS_3DParameters}(c). This definition has an advantage that the triple line is directly visible, and therefore it is easy to be measured in experiments. The second definition instead would better approximate the overall drop shape in the case of large ridges \cite{gunjan2020droplets}. In the limit of vanishing lubricant ridges, the two definitions coincide and the apparent contact angle can be expressed as an average of the effective lubricant-drop and lubricant-gas contact angles, weighted by the ratios of surface tensions:
	\begin{align}\label{AppAngle}
	\cos \theta_{app} & = - \cos \theta^{\rm eff}_{ld} \frac{\gamma_{ld}}{\gamma_{dg}} + \cos \theta^{\rm eff}_{lg}  \frac{\gamma_{lg}}{\gamma_{dg}} \nonumber \\
	& = - \cos \theta^{\rm eff}_{ld} \frac{\sin \theta_{g}}{\sin \theta_{l}} + \cos \theta^{\rm eff}_{lg} \frac{\sin \theta_{d}}{\sin \theta_{l}}.
	\end{align}
Unless specified otherwise, this equation defines the apparent angle used throughout the paper. It allows us to compare the wettability of droplets on LIS with those on homogeneous surfaces, employing $\theta_{app}$ for the former and $\theta_{dg}$ for the latter.

Throughout the paper, we non-dimensionalize the separation by the characteristic length scale of the deposited droplet, $s = \left(3 V_{d}/4 \pi \right)^{1/3}$; the lubricant pressure by a characteristic Laplace pressure for a droplet with surface tension $\gamma_{dg}$ and volume $V_d$, given by $\gamma_{dg}/s$; and the capillary force by a characteristic force, $2\pi \gamma_{dg} s$.
	
\subsection{Numerical Method}
We compute the liquid bridge morphologies in mechanical equilibrium by numerically minimising the free energy. To do this, we employ the public domain software Surface Evolver \cite{brakke1992surface,brakke_surface_1996}. Owing to the symmetry of the problem, we employ an effective 2D model with rotational symmetry \cite{semprebon2017apparent,de2008capillary,de2008effect,de2008enhancement} . 

The three fluid interfaces between drop, lubricant and gas are modelled by discrete segments joining in a point representing the drop-lubricant-gas triple line. The discrete segments are weighted appropriately by the relevant surface tensions. 

In this effective 2D model, each of the lubricant-drop and lubricant-gas interface also meets the substrate at a point corresponding to the substrate-lubricant-drop and substrate-lubricant-gas contact lines in the case of finite wettability. The relevant fluid-solid energy terms can then be calculated given the positions of the substrate-lubricant-drop and substrate-lubricant-gas contact lines. 

In the case of full wetting, there are no substrate-lubricant-drop and substrate-lubricant-gas contact lines. Instead, the fluid-solid energy is accounted by numerically integrating the disjoining pressure term. In this case we ensure that the lubricant layer has reached a constant thickness far away from the droplet in the case of full wetting. For this reason, we also apply a symmetric boundary condition with the external wall for the lubricant-gas interface.

Importantly, for all the results presented in this work, the drop volume is fixed by an integral constraint, while the lubricant exchange between the ridge and the surrounding reservoir is modelled by the $-\Delta P_{lg} V_{l}$ term as described in Eq.~\eqref{eq:FF}.

\subsection{Calculating the Forces and the Spring Constant}

	\begin{figure}
		\centering
		\includegraphics[width=0.75\linewidth]{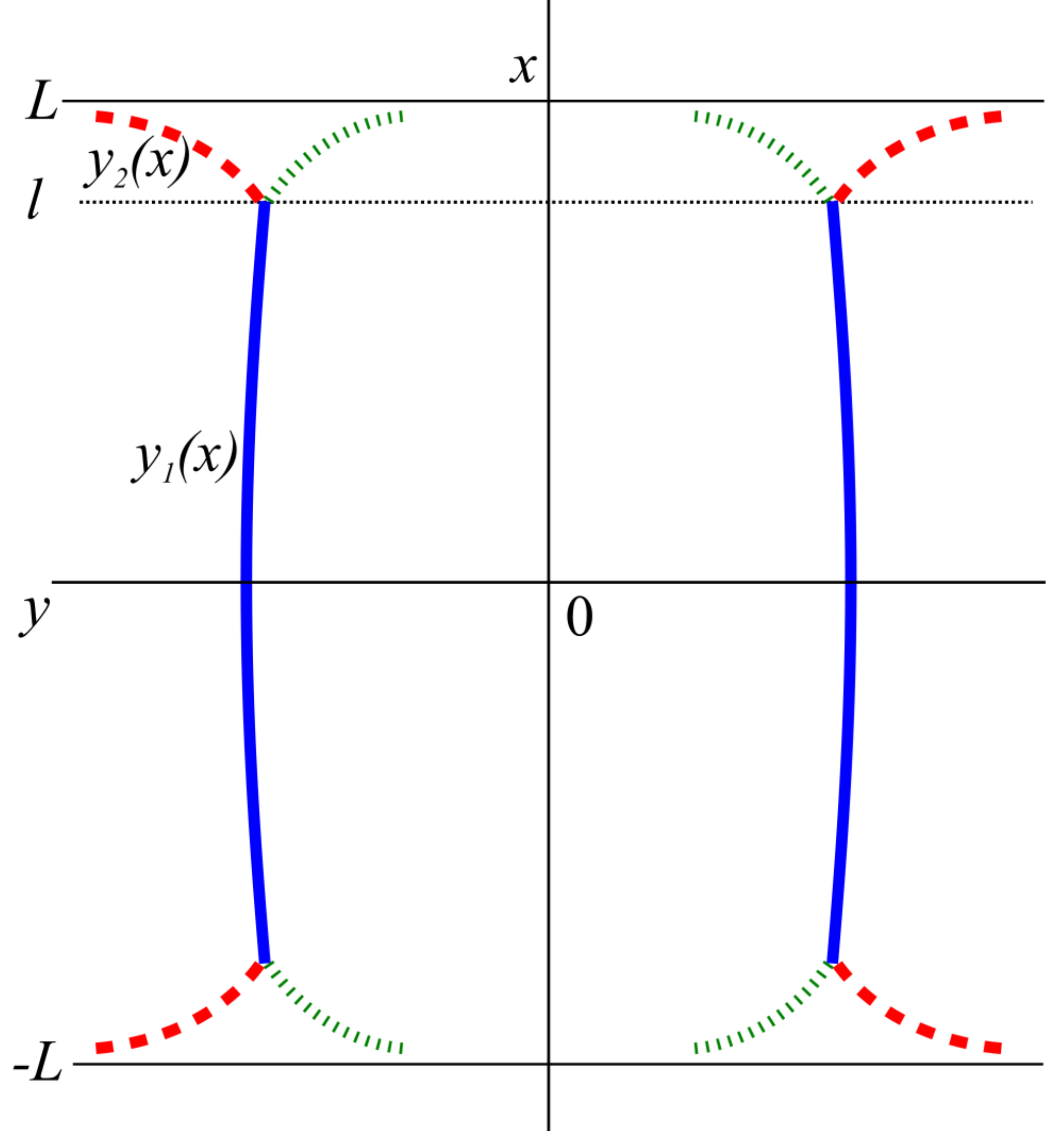}
		\caption{\label{TheoryProfile} Schematic of the capillary bridge profile. Here, $y_{1}(x)$ (solid line) is the drop-gas interface, $y_2(x)$ (dashed line) is the lubricant-gas interface, and $y_{3}(x)$ (dotted line) is the lubricant-drop interface.
		}
	\end{figure}
	
A natural measure of the ``stickiness'' of capillary bridges between LIS surfaces is given by the capillary force exerted by the bridges on the two substrates. This force can be directly calculated from the profile of a liquid bridge in mechanical equilibrium, as illustrated in Fig. \ref{TheoryProfile}. The detailed derivation of the expression for the force is reported in the Appendix. There are two natural locations to evaluate the capillary force. At the substrate, the capillary force is given by
	\begin{align} \label{eqn:force balance}
		F & = 2 \pi \left[\gamma_{lg} \frac{y_2 }{(1+{y_2^\prime}^2)^\frac{1}{2}} + \gamma_{ld} \frac{y_3}{(1+{y_3^\prime}^2)^\frac{1}{2}} \right]_{x=L} \nonumber\\
		& \quad - \pi \left[\Delta P_{dg} y_3^2 + \Delta P_{lg} (y_2^2 - y_3^2) \right]_{x=L}.
	\end{align}
	The first term can be interpreted as the surface tension forces of both the lubricant-gas and lubricant-droplet interfaces acting perpendicular to the substrate, while the second term corresponds to force contributions due to the droplet and lubricant pressures relative to the surrounding gas pressure. 
	
Alternatively,  the capillary force can be conveniently computed at the symmetry plane located at $x=0$ in the schematic diagram in Fig 2, in which $y_1^{\prime} = 0$. In this case, the force expression simplifies to 
\begin{eqnarray} \label{eqn:force balance2}
F &=&  2 \pi \left[\gamma_{dg}\frac{y_1}{(1+{y_1^\prime}^2)^\frac{1}{2}}\right]_{x=0} - \pi \left.\Delta P_{dg} y_1^2\right|_{x=0}, \nonumber \\ 
&=&  2 \pi \gamma_{dg}r_d - \pi \Delta P_{dg} r_d^2. 
\end{eqnarray}
Here $r_d$ is the radial distance of the droplet-gas interface. The correct implementation of these calculations has been validated by comparing them to numerical derivatives of the energy with respect to the plate separation.
	
A complementary measure of the ``stickiness'' is given by the spring constant of the bridge, $K_{eq}$, when the capillary force vanishes. To calculate $K_{eq}$, we can employ a simple linear relation between the calculated force and the displacement from the capillary bridge equilibrium distance. Following Hooke's Law,
\begin{align*}
F = -K_{eq} (H-H_{eq}),
\end{align*} 
where $H_{eq}$ is the capillary bridge separation when $F=0$. By fitting how $F$ varies with $H$ near the equilibrium distance, $K_{eq}$ can be obtained.
	
\section{Results: Study of capillary bridges between liquid infused surfaces} \label{results}

\subsection{Morphology Classes}

\begin{figure}
\centering
\includegraphics[width=\linewidth]{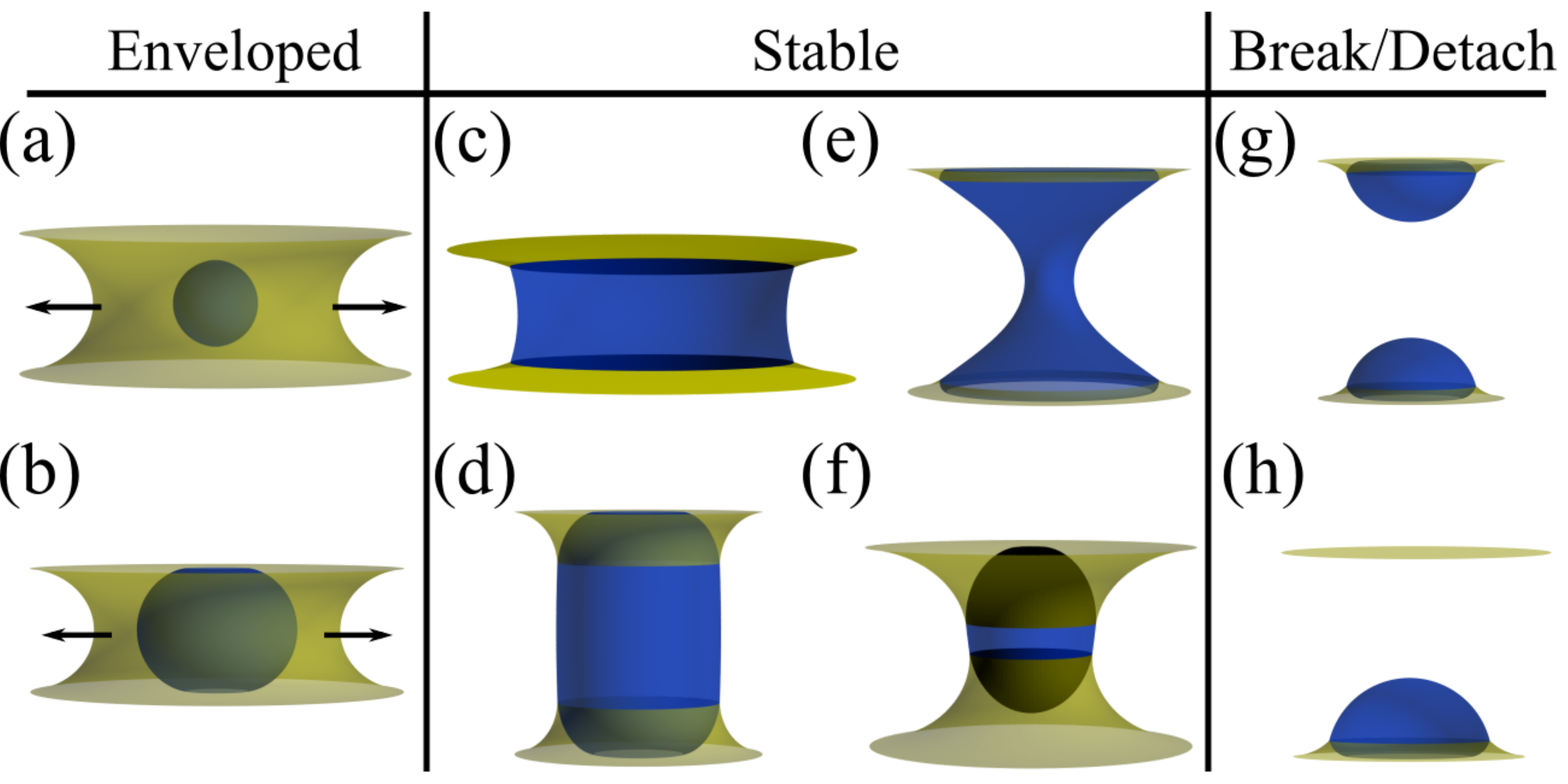}
\caption{Three broad categories of the energy minimization results. The top and bottom rows illustrate the typical morphologies for hydrophilic ($\theta_{app} < 90^\circ$) and hydrophobic ($\theta_{app} > 90^\circ$) capillary bridges, respectively. First, at small separation, the capillary bridge is unstable and the droplet is enveloped by the lubricant. Here, the droplet can be (a) detached from or (b) attached to the substrates. Second, stable capillary bridges are formed. For hydrophilic bridges, the droplet can be (c) detached from both substrates or (e) attached to both substrates; while for hydrophobic bridges, the droplet can be (d) attached to both substrates or (f) attached to only one substrate. Third, at large separation, the capillary bridges can become unstable by (g) breaking the droplet or (h) transferring to one of the substrates. \label{Morphologies}}
\end{figure}

\begin{figure*}
\centering
\includegraphics[width=0.9\textwidth]{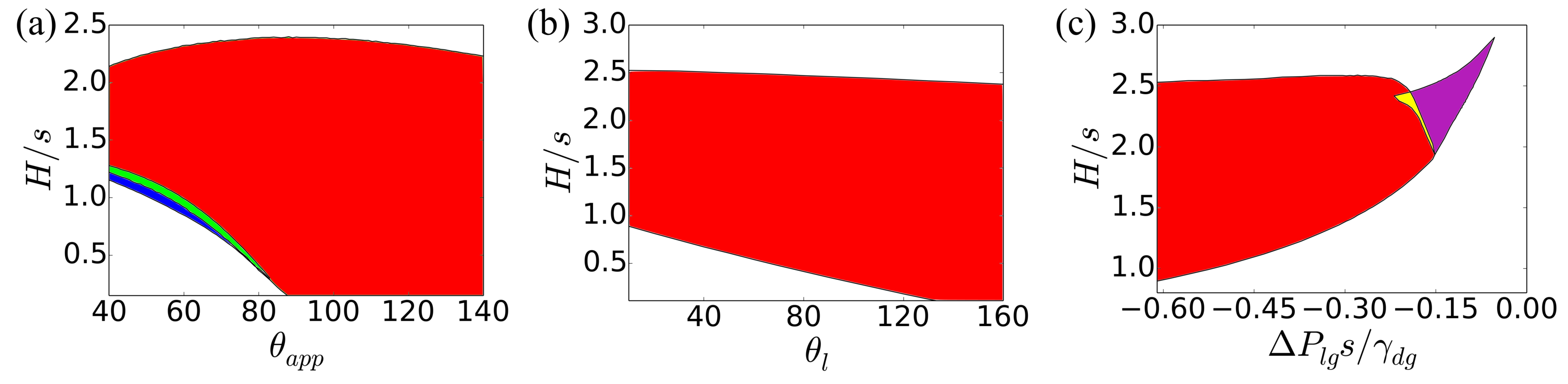}
\caption{\label{PhaseDiagrams} Morphology phase diagrams for a capillary bridge between two LIS.  In panel (a), we vary the apparent contact angle $\theta_{app}$ and normalised separation $H/s$, and fix the lubricant Neumann angle $\theta_{l} = 160^{\circ}$ and normalised lubricant pressure $\Delta P_{lg}s/\gamma_{dg} = -0.62$. In panel (b), we vary $H/s$ and $\theta_l$, keeping  $\theta_{app} = 100^{\circ}$ and $\Delta P_{lg}s/\gamma_{dg} = -0.62$. In panel (c), $H/s$ and $\Delta P_{lg}s/\gamma_{dg}$ are varied, with $\theta_{app}=100^{\circ}$ and $\theta_{l}=10^{\circ}$. In all panels, the wetting contact angles for the lubricant are set at $\theta_{ld}^{\rm eff}=\theta_{lg}^{\rm eff}=5^\circ$. The red region denotes stable morphologies when the droplet remains attached to both LIS substrates, blue when the droplet is detached from both substrates, and purple when the droplet is detached from one of the substrates. The green and yellow regions in panels (a) and (c) represent regions where multiple morphologies are possible.}
\end{figure*}

In this work we performed systematic numerical energy minimization varying the plate separation $H$, the apparent contact angle $\theta_{app}$, the lubricant Neumann angle $\theta_{l}$, and the lubricant pressure $\Delta P_{lg}$. We can broadly group the bridge morphologies on LIS into three categories. Referring to panels in Fig. \ref{Morphologies} these are: (a-b) unstable bridges due to envelopment instability, (c-f) stable bridges, and (g-h) unstable bridges due to capillary break up or detachment. Strictly speaking, the minimization routine fully converges only for morphologies labelled as stable. The unstable morphologies are depicted to illustrate the capillary bridges as the instabilities take place. Due to the large number of control parameters, we have computed morphology diagrams in the form of 2D slices of the parameter space, as depicted in Fig. \ref{PhaseDiagrams}. For each slice, all the remaining parameters are fixed. 

To start, let us vary the apparent angle  $\theta_{app}$ and plate separation $H/s$ in Fig. \ref{PhaseDiagrams}(a). Here we have fixed the wetting contact angles for the lubricant, $\theta_{ld}^{\rm eff}=\theta_{lg}^{\rm eff}=5^\circ$, the lubricant Neumann angle,  $\theta_{l} = 160^{\circ}$, and the normalised lubricant pressure, $\Delta P_{lg}s/\gamma_{dg} = -0.62$. Such lubricant pressure results in lubricant ridge that is small compared to the droplet size for the majority of the results shown in Fig. \ref{PhaseDiagrams}(a), as commonly observed in experiments \cite{smith2013droplet, schellenberger_direct_2015,mchale2019apparent}. However, the lubricant ridge can become comparable in size with the droplet at small separations.
	 
The largest, red colored region in Fig. \ref{PhaseDiagrams}(a) identifies the most common 
case of a stable liquid bridge in contact with both LIS substrates, as illustrated in Fig. \ref{Morphologies}(d-e). Moving to the bottom-left of the phase diagram by decreasing both the apparent angle and the plate separation, there also exists a morphology in which the droplet is detached from both substrates for hydrophilic capillary bridges ($\theta_{app} < 90^\circ$), and where lubricant menisci connect the droplet and the substrates. Here, as we reduce the plate separation, we find the Neumann triangle at the top of the lubricant ridge rotates such that the drop-gas interface becomes more aligned to the direction normal to the substrate. This in turn leads to a detachment of the lubricant-drop interface from the substrate. Such morphology is shown in Fig. \ref{Morphologies}(c) and corresponds to the blue region on the phase diagram in Fig. \ref{PhaseDiagrams}(a). The green region highlights the parameter space in which both morphologies are possible. 

Further decreasing the plate separation leads to an instability related to the coalescence of the two lubricant menisci. The lubricant floods the space in between the LIS substrates and envelopes the droplet. In this work, since we use the pressure ensemble for the lubricant, this continues indefinitely. However, in practice, especially in the limit of starved lubricant regime, it will do so until the lubricant from the drop surrounding is depleted. This envelopment instability is unique to LIS where the capillary bridge has two liquid components. Inspecting the data reported in Fig. \ref{PhaseDiagrams}(a), we further observe that the envelopment instability occurs at larger separation as we decrease the apparent angle. Typically, the envelopment occurs with the droplet detached from the surfaces for hydrophilic bridges ($\theta_{app} < 90^\circ$), as illustrated in Fig. \ref{Morphologies}(a); and with the droplet attached to the surfaces for hydrophobic bridges ($\theta_{app} > 90^\circ$), illustrated in Fig. \ref{Morphologies}(b).

The capillary bridges also become unstable at large plate separation. Similar to conventional one-component capillary bridges on non-infused surfaces, this can occur in two ways. The drop can break in two as in Fig. \ref{Morphologies}(g), typically observed for hydrophilic capillary bridges. Alternatively, the droplet may detach from one of the substrates, as illustrated in Fig. \ref{Morphologies}(h), commonly found for hydrophobic bridges. Since we ignore gravity in this work, the droplet can move to the top or bottom substrate with equal probability triggered by numerical noise during minimization. The detailed dynamics of these instabilities are beyond the scope of this work.
	 	 
Next, we study the role of the lubricant Neumann angle $\theta_{l}$, in particular to characterize the envelopment instability at small $H/s$. Fig. \ref{PhaseDiagrams}(b) shows a phase diagram as a function of $H/s$ and $\theta_{l}$. We now fix $\theta_{app} = 100^\circ$, while the other parameters are set as before in Fig. \ref{PhaseDiagrams} (a). It is worth noting that the other two Neumann angles are no longer independent variables when an apparent angle and one of the Neumann angles are defined. Here we employ the lubricant Neumann angle as control parameter, as it provides a good measure of the shape of the lubricant ridge. 

\begin{figure*}
\centering
\includegraphics[width=0.9\textwidth]{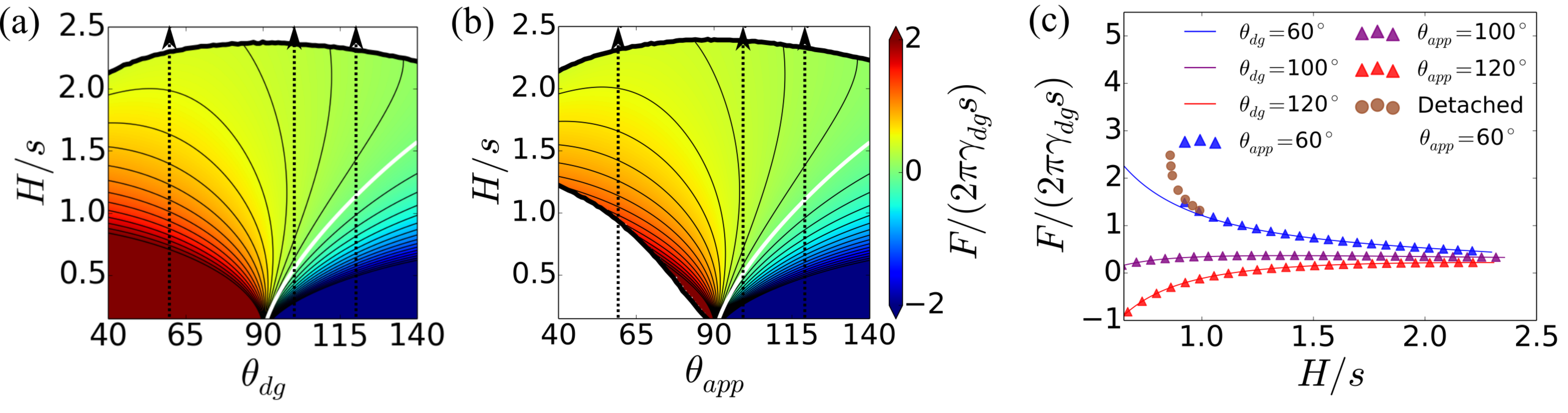}
\caption{\label{ForceContourAll} (a) and (b) Force contour plots as a function of contact angle ($\theta_{dg}$ or $\theta_{app}$) and substrate separation for smooth solid surfaces and LIS, respectively. For the latter, we have set $\theta_{l}=160^\circ$, $\theta_{ld}^{\rm eff}=\theta_{lg}^{\rm eff}=5^\circ$ and $\Delta P_{lg}s/\gamma_{dg} = -0.62$.  The white lines trace the separation for which the force is zero. (c) Capillary force as a function of the substrate separation for contact angles $\theta_{app} = 60^\circ$, $100^\circ$ and $120^\circ$. The LIS cases are represented by triangles and circles for when the droplet is attached and detached from the substrates, while those for smooth surfaces are represented by solid lines.}
\end{figure*}
	 
We clearly observe that the system is more prone to envelopment instability for smaller $\theta_{l}$. Considering the force balance at the droplet-lubricant-gas contact line, this is because smaller $\theta_{l}$ implies the droplet-gas surface tension becomes more dominant in magnitude over the droplet-lubricant and lubricant-gas surface tensions. As such, the system can lower its free energy by reducing any droplet-gas interface. In contrast, the detachment instability at large $H/s$ is relatively insensitive to $\theta_{l}$. 

In Fig. \ref{PhaseDiagrams}(a) we observe that the stability region for bridge morphologies where the drop is detached from both substrates, as depicted in Fig. \ref{Morphologies}(c), is limited to $\theta_{app} < 90^\circ$. We now investigate whether this morphology can occur also for $\theta_{app} > 90^\circ$ for a different combination of parameters. We find the critical parameter is the lubricant pressure $\Delta P_{lg}$. Thus, we present a morphology phase diagram as a function of separation $H/s$ and $\Delta P_{lg}s/\gamma_{dg}$ in Fig. \ref{PhaseDiagrams}(c). Here we fix $\theta_{ld}^{\rm eff}=\theta_{lg}^{\rm eff}=5^\circ$, $\theta_{app} = 100^\circ$, and $\theta_l = 10 ^\circ$. Interestingly, the morphology where the droplet is detached from both substrates is not found to be stable for hydrophobic bridges. Rather, we find a morphology where the droplet is only directly in contact with one of the substrates, as illustrated in Fig. \ref{Morphologies}(f), when we increase $\Delta P_{lg}s/\gamma_{dg}$ (becoming less negative).  Physically, increasing $\Delta P_{lg}s/\gamma_{dg}$ corresponds to reducing the energy costs of drawing lubricant from the reservoir, leading to lubricant meniscus size which is comparable to the droplet size. In the phase diagram of Fig. \ref{PhaseDiagrams}(c), this morphology is indicated by the purple region, bounded from below by the envelopment instability and from above by the droplet detachment instability. The two instability boundaries terminate in a cusp located at $\Delta P_{lg}s/\gamma_{dg} = -0.053$ and $H/s = 2.896$. It is also worth commenting that the morphology in Fig. \ref{Morphologies}(f) is observed for a wide range of $\theta_{l}$. Here we have fixed $\theta_l = 10^{\circ}$ because lower $\theta_l$ allows a broader range of plate separation in which the morphology is stable. The yellow region corresponds to parameter regime where the morphologies of Fig. \ref{Morphologies}(d) and (f) are both possible. 	 

\subsection{Capillary Forces}

Having discussed the spectrum of different liquid bridge morphologies between two LIS substrates, we now address the key question of how the forces exerted by stable capillary bridges are affected by the presence of the lubricant.	We will start by considering cases in which the lubricant ridge is small and flat, which we can achieve by setting large lubricant Neumann angle and low lubricant pressure. Here we use $\theta_l = 160^\circ$, and $\Delta P_{lg}s/\gamma_{dg} = -0.62$. Fig. \ref{ForceContourAll}(a) and (b) show the (normalised) force contour plots as a function of the (apparent) contact angle and plate separation for a capillary bridge on smooth surfaces and LIS, respectively. For the latter, we focus on cases where the droplet is directly in contact with both LIS substrates. The parameter space explored here is the same as in the morphology diagram in Fig. \ref{PhaseDiagrams}(a).

 In this limit in which the lubricant ridge is small and flat, the capillary forces obtained are quantitatively very similar for LIS and for smooth solid surfaces. For hydrophilic capillary bridges ($\theta_{app}< 90^\circ$), the force is always attractive and monotonically decreases as one increases $H/s$; while for  hydrophobic capillary bridges ($\theta_{app}> 90^\circ$), the force is repulsive at short separation and attractive at large separation. Following Eqs.~\ref{eqn:force balance} and \ref{eqn:force balance2}, this is because while the surface tension contributions are always attractive, the pressure contributions can be either attractive or repulsive. The contour of zero force, corresponding to the equilibrium separation distance, $H_{eq}$, is marked by the white lines in Fig. \ref{ForceContourAll} (a) and (b). From these results, we can also conclude that the apparent contact angle is a suitable control parameter to compare capillary bridges on conventional smooth, non-infused surfaces and LIS in the limit of small lubricant ridges.
 	
The main difference between one-component liquid bridges on non-infused surfaces and two-component liquid bridges on LIS, as shown in Fig. \ref{ForceContourAll}, is found at small $H/s$ for $\theta_{app}< 90^\circ$, as we approach the parameter space in which the detached droplet and enveloped morphologies in Figs. \ref{Morphologies}(a) and (c) are preferred. To better illustrate this point, Fig. \ref{ForceContourAll}(c) plots the capillary force $F$ as a function of separation $H$ for three contact angles, $60^\circ$, $100^\circ$ and $120^\circ$. The LIS cases are shown with triangular and circular markers for when the droplet is attached and detached from the substrates, while those for one-component capillary bridge on smooth solid surfaces are shown with solid lines. As illustrated in Fig. \ref{ForceContourAll}(c), we see the force deviates strongly at small $H/s$ for $\theta_{app}=60^{\circ}$. This is due to significant amount of lubricant being drawn from the reservoir, increasing the attractive force of the capillary bridge.	

This deviation illustrates the importance of the lubricant ridge in the resulting capillary force. Since the shape and size of the ridge depends on the Neumann and lubricant contact angles, as well as the lubricant pressure, we further investigate how these parameters affect the strength of the capillary adhesion of the droplet on LIS. To study these effects we have chosen to employ an apparent angle of $\theta_{app}=100^\circ$ as an exemplary case. Such contact angle value is quite typical in experiments \cite{smith2013droplet, schellenberger_direct_2015}, though of course the apparent contact angle can be varied depending on the materials involved.

	\begin{figure*}
			\centering
			\includegraphics[width=0.85\textwidth]{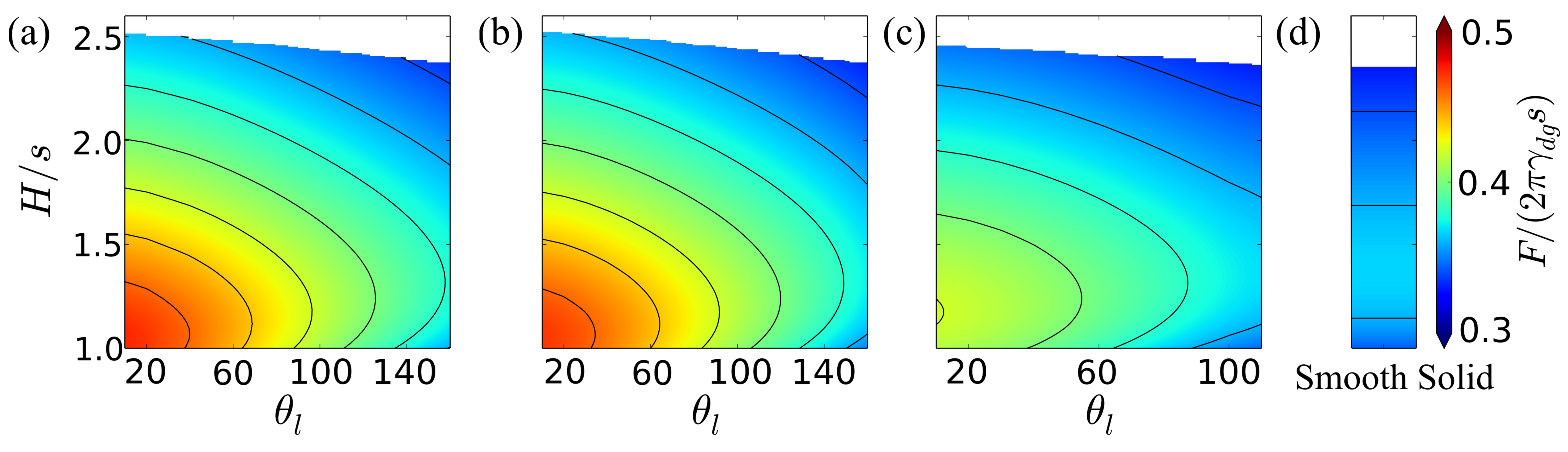}
			\caption{\label{CA_Wetting} (a-c) Force contour plots as a function of the lubricant Neumann angle $\theta_{l}$ and normalised separation $H/s$.  In panels (a-c), we have respectively set $\theta^{\rm eff}_{ld}$, $\theta^{\rm eff}_{lg}=0^\circ$, $5^\circ$, and $30^\circ$. In all cases, we have $\theta_{app} = 100^\circ$ and $\Delta P_{lg}s/\gamma_{dg} = -0.62$. For the complete wetting case, we have also chosen a Hamaker constant of $B/\gamma_{dg}s^2= 2.6 \times 10^{-8}$. (d) Force contour plot as a function of $H/s$ for a one-component capillary bridge with $\theta_{dg} = 100^\circ$.}
	\end{figure*}
	
	The effect of altering the wetting contact angles for the lubricant $\theta_{lg}^{\rm eff}$, $\theta_{ld}^{\rm eff} $ and the Neumann angle $\theta_{l}$ at fixed apparent angle is shown in Fig. \ref{CA_Wetting}.  Fig. \ref{CA_Wetting}(a)-(c) show contour plots of the (normalised) force as function of $\theta_{l}$ and $H/s$ for different lubricant contact angles.  In all three plots we have chosen $\Delta P_{lg}s/\gamma_{dg} = -0.62$. For comparison, we also show the force contour plot for one-component capillary bridge with $\theta_{dg} = 100^\circ$ in Fig. \ref{CA_Wetting}(d).
	
	The case with completely wetting lubricant is shown in Fig. \ref{CA_Wetting}(a). Here we need to include the physics of disjoining pressure, and we have chosen a Hamaker constant of $B/\gamma_{dg}s^2= 2.6 \times 10^{-8}$ to ensure the thickness of the lubricant layer is several orders of magnitude, $O(10^3)$, smaller than the droplet size $V_{d}^{1/3}$, as is typical in experiments \cite{daniel2017oleoplaning,kreder2018film}. We have also studied cases with larger disjoining pressure (data not shown), and the only difference is that the lubricant layer thickness offsets the value of $H/s$ when comparing the capillary forces. From the plot in Fig. \ref{CA_Wetting}(a), the maximum capillary force and separation distance increases with decreasing $\theta_{l}$. 
	
	The same tendency in the variation of $\theta_{l}$ is observed in Fig. \ref{CA_Wetting}(b) and (c), where we now employ partially wetting lubricants with $\theta_{lg}^{\rm eff} = \theta_{ld}^{\rm eff} = 5^\circ$ and $30^\circ$. We note that, for panel (c), there is no available solution for $\theta_l > 120^{\circ}$ because the three angles forming the lubricant ridge must sum to less than $180^{\circ}$. Comparing the results in Fig. \ref{CA_Wetting}(a)-(d), it is also clear that the capillary adhesion is stronger on LIS than for one-component capillary bridge on smooth surfaces, and that this ``stickiness" is amplified for more wetting lubricants. We find the capillary force on LIS can be up to $40\%$ higher than that on smooth solid surfaces. Taking the results in Fig. \ref{CA_Wetting} together, the shape of the lubricant ridge is an important factor to manipulate the strength of the capillary force. 

	\begin{figure*}
		\centering
		\includegraphics[width=0.9\textwidth]{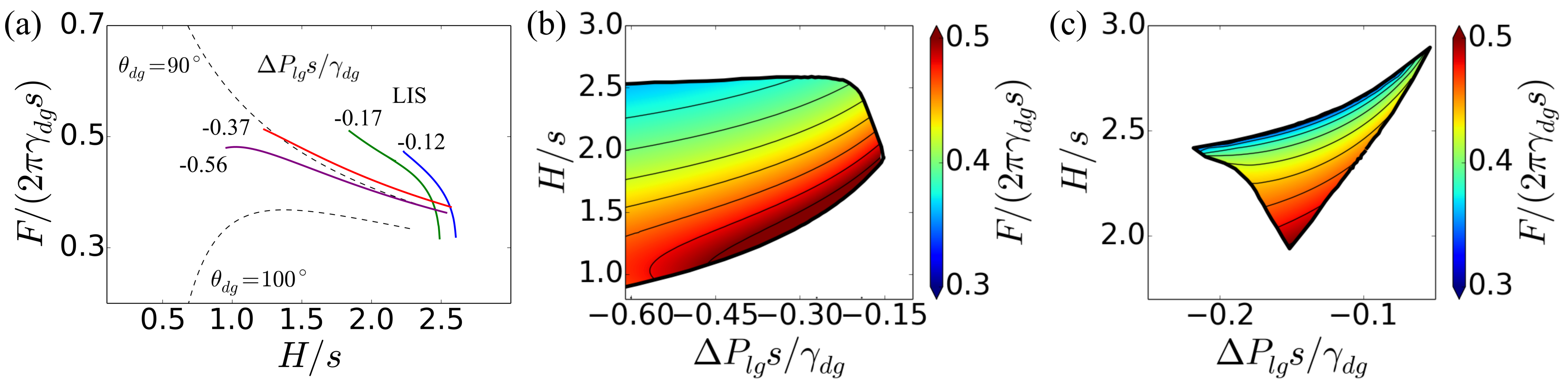}
		\caption{(a) Normalised capillary force as a function of normalised separation for several values of normalised lubricant pressure and an apparent angle of $\theta_{app} = 100^\circ$. For comparison, the force curves for one-component capillary bridge is also shown for $\theta_{dg}=90^{\circ}$, $100^{\circ}$. (b-c) Force contour plots as a function of the normalised separation and normalised lubricant pressure when the droplet is connected to (b) both LIS substrates and (c) only one of the LIS substrates. Here we have used $\theta_{l}=10^\circ$, and $\theta^{\rm eff}_{ld}, \theta^{\rm eff}_{lg}=5^\circ$.		
		\label{ForcePoil} }
	\end{figure*}
		
	Next, we address the issue of the size of the lubricant ridge, which can be tuned by varying the lubricant pressure. Fig. \ref{ForcePoil} (a) illustrates the case for a droplet with $\theta_{app} = 100^{\circ}$, $\theta_{l}=10^{\circ}$, $\theta_{ld}^{\rm eff} = \theta_{lg}^{\rm eff} = 5^{\circ}$, while $\Delta P_{lg}s/\gamma_{dg}$ is varied. Here, the force curves only exist over a certain range of $H/s$, bounded by the envelopment instability from below and the detachment instability from above, as discussed in the previous section. 
	
	It is known that the apparent angle as defined in Eq. \ref{AppAngle} is valid for the limit of vanishing ridge. As we make $\Delta P_{lg}s/\gamma_{dg}$ less negative, the lubricant ridge increases in size in comparison to the droplet size, and this can lead to a reduction in the measured geometric apparent angle \cite{semprebon2017apparent,sadullah2018drop}. For hydrophobic cases, this angle will always be bounded from below by $90^{\circ}$. Therefore, to test whether the increase in the capillary force is simply due to changes in the effective apparent angle, we also plot force curves for one-component capillary bridges in Fig. \ref{ForcePoil}(a) for contact angles $\theta_{dg}=90^{\circ}$, $100^{\circ}$, to allow for easy comparison. We find that we can even exceed the force from the $90^{\circ}$ case for certain separation distances. This implies that the apparent angle alone is insufficient to fully describe the increase in capillary force.  

	Fig. \ref{ForcePoil}(b) and (c) summarise the force contour plots as a function of $\Delta P_{lg}s/\gamma_{dg}$ and $H$ for the stable morphologies where the droplet is directly attached to both LIS substrates and only one of the substrates respectively.  In both scenarios, the capillary force increases with increasing $\Delta P_{lg}s/\gamma_{dg}$ at constant $H/s$, and decreases with $H/s$ at constant $\Delta P_{lg}s/\gamma_{dg}$.  However, comparing the two contour plots, the capillary force is affected differently. This is most clearly illustrated by the force curves in Fig. \ref{ForcePoil}(a). The cases of $\Delta P_{lg}s/\gamma_{dg} = -0.17$ and $-0.12$ have a different shape, with a greater dip in force near their maximum separation where droplet detachment occurs. This is because these cases correspond to morphology in Fig. \ref{Morphologies}(f), where the droplet is only directly connected to one of the LIS substrates. Here, the capillary force is dominated by the drawn lubricant. In contrast, the cases of $\Delta P_{lg}s/\gamma_{dg} = -0.56$ and $-0.37$ correspond to the morphology in Fig. \ref{Morphologies}(d), where the droplet dominates the capillary force response. 
	
\subsection{Equilibrium Separation and Spring Constant}

To further corroborate our findings on the adhesion of a liquid droplet on LIS, in this subsection we will consider how the equilibrium separation, $H_{eq}$, and spring constant, $K_{eq}$, of the capillary bridge compare to those for one-component capillary bridge on non-infused, smooth solid surfaces. 
		
	\begin{figure} [t]
	\centering
	\includegraphics[width=1.0\linewidth]{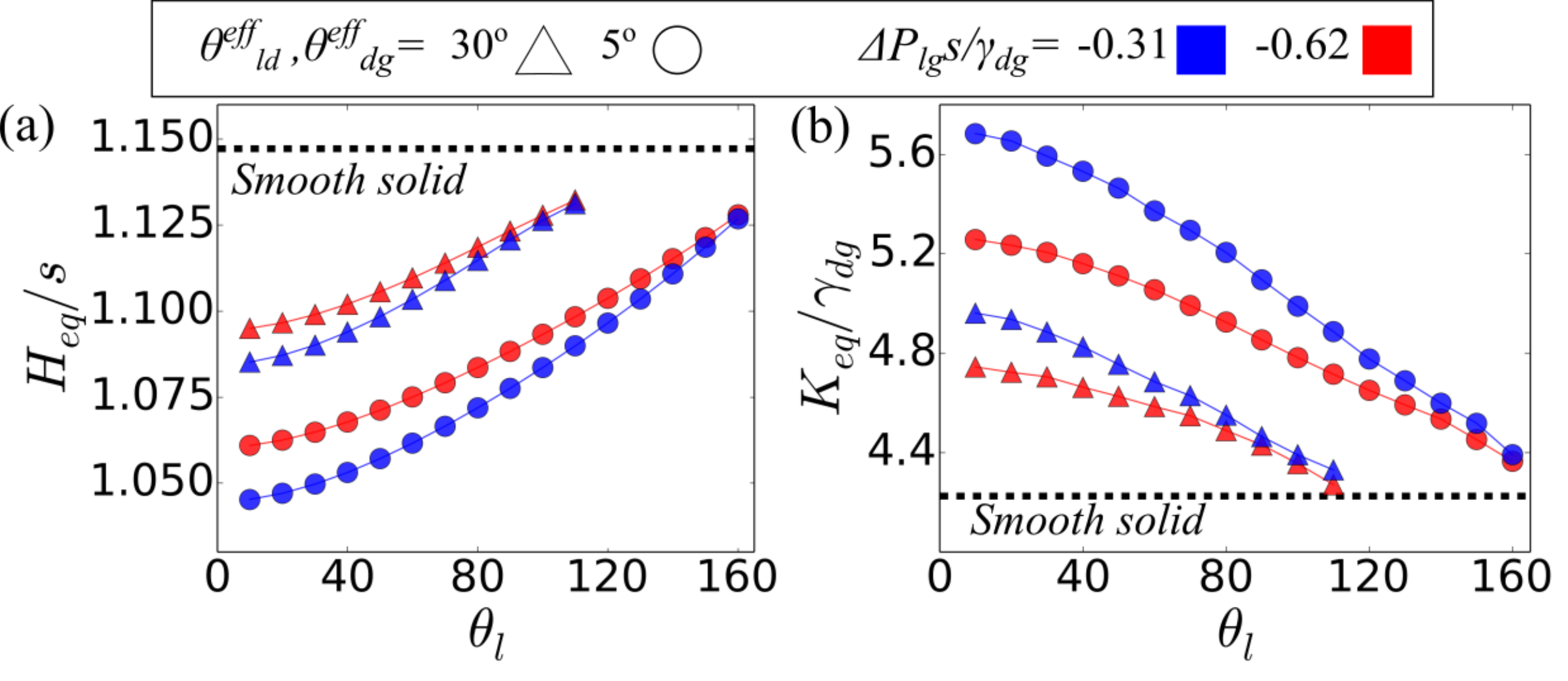}
	\caption{In both plots, the circles and triangles correspond to $\theta^{\rm eff}_{ld}, \theta^{\rm eff}_{lg} = 5^\circ$ and $30^\circ$, respectively; while the straight and dashed lines represent $\Delta P_{lg}s/\gamma_{dg} = -0.31$ and $-0.62$. The apparent contact angle is set at $\theta_{app}=120^\circ$. (a) Equilibrium separation $H_{eq}$ as a function of the lubricant Neumann angle. (b) Effective spring constant, $K_{eq}$ as a function of the lubricant Neumann angle. The results for one-component capillary bridge on smooth surfaces are shown by the dotted lines. 	
	\label{ForceSpringCompare}}
	\end{figure} 
	
	We first note that, for one-component capillary bridge on non-infused surfaces, there is always an equilibrium separation in which the capillary force $F = 0$ for $\theta_{dg} > 90^\circ$. In contrast, this is not the case on LIS. In particular, the envelopment instability often kicks in before the capillary bridge can reach its equilibrium separation distance for small but negative value of $\Delta P_{lg}s/\gamma_{dg}$. For $\theta_{app} = 100^\circ$, this instability significantly limits the range of parameters where the capillary bridge can reach $F = 0$. As such, in this subsection, we have chosen to focus on results with $\theta_{app} = 120^\circ$. The qualitative behaviour is the same for different apparent angles.
		
	Fig. \ref{ForceSpringCompare} summarises our findings. In panel (a), we first study how the equilibrium separation is affected by the Neumann angle, $\theta_l$, the wetting contact angles for the lubricant, $\theta_{lg}^{\rm eff}$, $\theta_{ld}^{\rm eff}$, and the lubricant pressure, $\Delta P_{lg}s/\gamma_{dg}$. For comparison, the one-component liquid bridge case is also shown as the black dotted line. Overall, we find the equilibrium separation is slightly smaller for LIS than for a one-component bridge. From the four curves for $H_{eq}$, we also observe that we approach the one-component liquid bridge case as we increase $\theta_l$ and make $\Delta P_{lg}s/\gamma_{dg}$ more negative. This is consistent with our observation in the previous subsection that the lubricant ridge morphology impacts the capillary force, so that as the ridge vanishes, LIS are equivalent to smooth surfaces in this limit.
		
	We also study the effective spring constant, $K_{eq}$, around the equilibrium separation in Fig. \ref{ForceSpringCompare}(b). $K_{eq}$ is always greater than that for the one-component liquid bridge implying that, not only capillary forces are larger on LIS, but also the resulting capillary bridges are stiffer due to the lubricant drawn from the LIS substrates. 

	\section{Conclusion}
	
	In summary we have studied two-component capillary bridges formed when a liquid droplet is sandwiched between two LIS. The lubricant ridge morphology was shown to be important in determining the capillary force, the maximum substrate separation, and the effective spring constant, which are all enhanced by lowering the lubricant-gas pressure difference $\Delta P_{og}$, the lubricant Neumann angle $\theta_{l}$, and the lubricant contact angles $\theta^{\rm eff}_{ld}, \theta^{\rm eff}_{lg}$ relative to the drop and gas phases. The parameter $\Delta P_{lg}$ affected the overall size of the lubricant ridge, $\theta_{l}$ changed the shape of the ridge, and $\theta^{\rm eff}_{ld}, \theta^{\rm eff}_{lg}$ tuned the adhesion between the lubricant and the solid surface. Varying these parameters we found that, for a given drop apparent contact angle, a capillary force could be achieved on LIS which was up to 40\% higher than that for the one-component case. Conversely, in the limit of a vanishing lubricant ridge, the properties of the two-component capillary bridge tended to those of a one-component bridge between smooth, solid substrates.

	In contrast to the one-component capillary bridge, a two-component capillary bridge also had a richer set of possible morphologies. Due to the presence of the lubricant, the capillary bridge could be stable with the liquid droplet directly in contact with two, one or none of the LIS substrates. At small separation, we have further identified a new envelopment instability, where the lubricant fills the space in between the two substrates. 
	
	While the facts that LIS have low contact angle hysteresis and are thus ``slippery'' parallel to the surface have been extensively discussed in the literature, here we point out that, at the same time, they are potentially ``sticky'' perpendicular to the surface. This suggests LIS are a unique class of liquid repellent surfaces. For instance, on superhydrophobic surfaces, liquid droplets are suspended on top of the surface textures in the Cassie-Baxter state, and as a result, they move easily both parallel and perpendicular to the surface \cite{roach2008progess,quere2008wetting,kusumaatmaja2007modeling}. In the Wenzel state, liquid drops penetrate the surface corrugations leading to both large contact angle hysteresis and strong adhesion \cite{roach2008progess,quere2008wetting,kusumaatmaja2007modeling}, and so they are ``sticky'' in both directions. 
		
	There are a number of directions for future work. In this work we have assumed the lubricant reservoir to be virtually infinite. However,	 LIS substrates can be prepared so that the lubricant layer has minimal thickness. In this case, the amount of lubricant which can be gathered will be limited, and this can affect the resulting liquid bridge morphologies. Another avenue is to extend our study to include cases where the droplet is encapsulated by a thin film of lubricant. This occurs when the wetting parameter, $S_{dl} = \gamma_{dg} - \gamma_{ld} - \gamma_{lg} < 0$. It is also interesting to consider potential applications exploiting the ``sticky''-``slippery'' behaviour of LIS, such as for microfluidic devices, fog harvesting and heat transfer. For instance, Launay {\it et al.} recently showed that LIS can capture droplets against gravity while allowing droplet transport via self-propulsion driven by topographical gradients \cite{launay2020self}. For these problems, it is important that we not only consider the static liquid bridge morphologies, but also the rich dynamics of the droplet deformation and lubricant flow. 
	
{\it{Acknowledgments}} - We thank Yonas Gizaw for useful discussions. ACMS is supported by EPSRC's Centre for Doctoral Training in Soft Matter and Functional Interfaces (EP/L015536/1; ACMS). HK and JRP acknowledge funding from Procter \& Gamble. CS acknowledges support from Northumbria University through the Vice-Chancellor's Fellowship Programme and EPSRC (EP/S036857/1) for funding.

	\bibliographystyle{achemso}
	\bibliography{Ref}

	\appendix
	\section{Thermodynamic Derivation of the Force}
	
	The capillary force exerted by the liquid bridge can be directly calculated from the equilibrium profile.  To derive the force, consider variation of the free energy of the system induced by varying the plate separation, $H \rightarrow H + \delta H$, following a similar approach carried out for one-component liquid bridges \cite{kusumaatmaja2010equilibrium,carter_forces_1988,brinkmann_wetting_2002}. Since we are considering symmetric surfaces, it is mathematically more convenient to set the origin of our coordinate system to be at the middle of the liquid bridge.  If we define the $y_{i}(x)$ to represent each interface's radial distance from the axial centre of the bridge, then the surfaces are located at $\pm H/2$, as defined in Fig \ref{TheoryProfile}.  As such, exploiting symmetry, we only need to consider the profile curves from $x=0$ to $x=L = H/2$. Under this variation, the position of the drop-lubricant-gas triple contact line varies from $l \rightarrow l + \delta l$, and correspondingly the profile curves also vary as $y_i(x) \rightarrow \bar{y}_i(x) = y_i(x)+g_i(x)$. 
	
The free energy upon extending the plate separation by $\delta H = 2\delta L$ is given by
	\begin{align}
	& \frac{E_{\rm LIS}(L+\delta L)}{2 \pi} \nonumber\\
	& = \int_{0}^{l+\delta l} \gamma_{dg} \bar{y}_{1}(x)(1+({\bar{y}_{1}^\prime})^2(x))^{1/2} - \frac{\Delta P_{dg}}{2} \bar{y}_{1}^2(x) \text{ d}x \nonumber\\
	& +\int_{l+\delta l}^{L+\delta L} \gamma_{lg} \bar{y}_{2}(x)(1+({\bar{y}_{2}^\prime})^2(x))^{1/2} - \frac{\Delta P_{lg}}{2}\bar{y}_2^2(x) \text{ d}x \nonumber\\
	& + \int_{l+\delta l}^{L+\delta L} \gamma_{ld} \bar{y}_{3}(x)(1+({\bar{y}_{3}^\prime})^2(x))^{1/2} - \frac{\Delta P_{dl}}{2} \bar{y}_3^2(x)\text{ d}x \nonumber\\
	& + \gamma_{ld}\cos\theta^{\rm eff}_{ld}\bar{y}_3^2(L+\delta L) - \gamma_{lg} \cos\theta^{\rm eff}_{lg}\bar{y}_2^2(L+\delta L).
	\end{align}
The first variation of the free energy leads to the well-known Euler-Lagrange or shape equation for each of the fluid interface, $y_{i}(x)$, as derived in \cite{kusumaatmaja2010equilibrium,carter_forces_1988,brinkmann_wetting_2002}. In addition, it leads to boundary conditions at $x = l$ and $x = L$. For the former, considering the variation in $\delta l$ gives us
	\begin{align}
	&\left[ \frac{\gamma_{dg}}{(1+({y_{1}^\prime})^2)^{1/2}} - \frac{\gamma_{lg}}{(1+({y_{2}^\prime})^2)^{1/2}} - \frac{\gamma_{ld}}{(1+({y_{3}^\prime})^2)^{1/2}} \right] y(l) \nonumber \\
	& - \left[\frac{\Delta P_{dg}}{2} - \frac{\Delta P_{lg}}{2} -  \frac{\Delta P_{dl}}{2} \right] y(l)^2 = 0.
	\end{align}
where we have used $y(l)=y_i(l)$. The second term in the above equation is identically zero, since the pressure terms cancel one another. Furthermore, the square bracket in the first term corresponds to the force balance of the surface tensions projected in the $x$-direction. Similarly, the variation in $\delta y(l)$ results in
	\begin{equation}
	\left[ \frac{\gamma_{dg} y_{1}^\prime}{(1+({y_{1}^\prime})^2)^{1/2}} - \frac{\gamma_{lg} y_{2}^\prime}{(1+({y_{2}^\prime})^2)^{1/2}} - \frac{\gamma_{ld} y_{3}^\prime}{(1+({y_{3}^\prime})^2)^{1/2}} \right] = 0,
	\end{equation}
	which is the force balance of the surface tensions projected in the $y$-direction. Note that for these derivations we have used the geometrical relation $g_i(l+\delta l) = \delta y_i (l)- y_i^{\prime}(l) \delta l$.
	
	We can do a similar analysis for the boundary condition at $x = L$. Considering the variation in $\delta y_2(L)$ and $\delta y_3(L)$, we recover the expected contact angle equations for the lubricant-gas and lubricant droplet interfaces,
	\begin{eqnarray}
	\left[\gamma_{lg} \frac{y_2 {y_2^{\prime}}}{(1+{y_2^{\prime}}^2)^{\frac{1}{2}}} \right]_{x=L}  - \gamma_{lg} \cos\theta_{lg}^{\rm eff} y_2(L) = 0, \\
	\left[\gamma_{ld} \frac{y_3 {y_3^{\prime}}}{(1+{y_3^{\prime}}^2)^{\frac{1}{2}}} \right]_{x=L} + \gamma_{ld} \cos\theta_{ld}^{\rm eff} y_3(L) = 0.
	\end{eqnarray}
	Finally, the variation in $\delta L$ leads to the equation for the force exerted by the capillary bridge, given by 
	\begin{align} \label{eqn:force balance_2}
	F & = \left.\frac{ \delta(2 E)}{\delta(2 L)}\right|_{x=L} \nonumber \\ 
	& = 2 \pi \left[\gamma_{lg} \frac{y_2 }{(1+{y_2^\prime}^2)^\frac{1}{2}} + \gamma_{ld} \frac{y_3}{(1+{y_3^\prime}^2)^\frac{1}{2}} \right]_{x=L} \nonumber\\
	& \quad - \pi \left[\Delta P_{dg} y_3^2 + \Delta P_{lg} (y_2^2 - y_3^2) \right]_{x=L}
	\end{align}
	
	At equilibrium, the force is constant everywhere along the capillary bridge. Therefore, it is often convenient to consider the force at the mid-point of the liquid bridge, where $x=0$ and $y_1^{\prime} = 0$. In this case, the force expression simplifies to 
	\begin{eqnarray} 
	F &=&  2 \pi \left[\gamma_{dg}\frac{y_1}{(1+{y_1^\prime}^2)^\frac{1}{2}}\right]_{x=0} - \pi \left[ \Delta P_{dg} y_1^2\right]_{x=0}, \nonumber \\ 
	&=&  2 \pi \gamma_{dg}r_d - \pi \Delta P_{dg} r_d^2. 
	\end{eqnarray}
where $r_d$ is the radial distance of the drop-gas interface at the middle of the bridge as defined as in Fig \ref{LIS_3DParameters}. 

\end{document}